\documentclass[10pt,conference]{IEEEtran}

\usepackage{booktabs}
\usepackage{graphicx}
\usepackage{textcomp}
\usepackage{xcolor}
\usepackage{xspace}
\usepackage{url}
\usepackage{bm}
\usepackage{tabularx}
\usepackage{color}
\usepackage{colortbl}
\usepackage[switch]{lineno}
\usepackage{subcaption}
\usepackage{caption}
\usepackage{tcolorbox}
\usepackage{bbding}
\usepackage{multicol}
\usepackage{amsmath}
\usepackage{hyperref} 
\usepackage{url} 
\usepackage{float} 
\usepackage{subcaption}

\usepackage{braket}

\usepackage{subcaption}

\definecolor{light-gray}{gray}{0.8}
\definecolor{light-red}{RGB}{255,155,155}
\definecolor{light-blue}{RGB}{0,155,255}
\definecolor{codegreen}{rgb}{0,0.6,0}
\definecolor{codegray}{rgb}{0.5,0.5,0.5}
\definecolor{codepurple}{rgb}{0.58,0,0.82}
\definecolor{backcolour}{rgb}{0.95,0.95,0.92}

\newcommand{\bench}[1]{#1\xspace}

\newcommand{\acrvtwo}{AutoCodeRover-v2\xspace}
\newcommand{\swebench}{SWE-bench\xspace}
\newcommand{\swebenchlite}{SWE-bench Lite\xspace}

\newcommand{\swelite}{\bench{SWE-bench Lite}}

\definecolor{sblue}{rgb}{0.36, 0.54, 0.66}

\definecolor{fig4green}{RGB}{151, 208, 119}
\definecolor{fig4blue}{RGB}{169, 196, 235}

\usepackage[colorinlistoftodos,prependcaption,textsize=tiny]{todonotes}

\usepackage{enumerate}
\usepackage[shortlabels]{enumitem}

\usepackage{tcolorbox}

\usepackage{multirow}
\usepackage{pbox}
\usepackage{tabularx}

\usepackage{listings}
\newcommand{\lstbg}[3][0pt]{{\fboxsep#1\colorbox{#2}{\strut #3}}}
\lstdefinestyle{mystyle}
{
    language = Python,
    basicstyle = {\ttfamily \color{main-color}},
    keywordstyle = {\color{blue}},
    keywordstyle = [2]{\color{blue}},
    keywordstyle = [3]{\color{yellow}},
    keywordstyle = [4]{\color{teal}},
    morekeywords = [3]{<<, >>},
    morekeywords = [4]{++},
    basicstyle=\ttfamily\footnotesize,
    commentstyle=\color{gray}\ttfamily,
    morecomment=[f][\lstbg{red!20}]-,
    morecomment=[f][\lstbg{green!20}]+,
    morecomment=[f][\lstbg{yellow!20}]++,
    morecomment=[f][\lstbg{yellow!20}]--,
    morecomment=[f][\textit]{@@},
    texcl=false
}

\lstdefinestyle{prompt_style}
{
    language = {},
    keywordstyle = {\color{blue}},
    keywordstyle = [2]{\color{blue}},
    keywordstyle = [3]{\color{yellow}},
    keywordstyle = [4]{\color{teal}},
    morekeywords = [3]{<<, >>},
    morekeywords = [4]{++},
    basicstyle=\ttfamily\footnotesize,
    commentstyle=\color{gray}\ttfamily,
    morecomment=[f][\lstbg{red!20}]-,
    morecomment=[f][\lstbg{green!20}]+,
    morecomment=[f][\lstbg{yellow!20}]++,
    morecomment=[f][\lstbg{yellow!20}]--,
    morecomment=[f][\textit]{@@},
    texcl=false,
    numbers=none,
    breakindent=0pt
}

\usepackage{etoolbox,environ}
\newtoggle{isArxivVersion}

\definecolor{fpbackcolor}{RGB}{242,242,242}
\definecolor{diffrem}{RGB}{202, 45, 49}
\definecolor{diffincl}{RGB}{0, 135, 90}
\definecolor{codepink}{RGB}{237, 2, 140}

\newboolean{showcomments}
\setboolean{showcomments}{true} 
\ifthenelse{\boolean{showcomments}}{
  \newcommand{\nbc}[3]{
    {\textcolor{#3}{\small{\bfseries{#1:\ }}\textit{#2}}}}
}{
  \newcommand{\nbc}[3]{}
}

\usepackage{amsfonts}

\usepackage{braket}

\usepackage{pifont}

\usepackage{xcolor}
\usepackage{soul}

\newcommand{\code}[1]{\texttt{#1}}

\newcommand{\toolname}{SpecRover\xspace}

\newcommand{\artifacturl}{\url{https://zenodo.org/doi/10.5281/zenodo.13161650}\xspace}

\newcommand{\new}[1]{{\color{black}{#1}\xspace}}

\IEEEoverridecommandlockouts
\usepackage{cite}
\usepackage{amsmath,amssymb,amsfonts}
\usepackage{algorithmic}
\usepackage{graphicx}
\usepackage{textcomp}
\usepackage{xcolor}
\def\BibTeX{{\rm B\kern-.05em{\sc i\kern-.025em b}\kern-.08em
    T\kern-.1667em\lower.7ex\hbox{E}\kern-.125emX}}

\begin{document}

\title{\toolname: Code Intent Extraction via LLMs
\thanks{\textsuperscript{*}Joint first authors, ordered alphabetically.}
}

\author{Haifeng Ruan\textsuperscript{*} \hspace*{0.2in} Yuntong Zhang\textsuperscript{*} \hspace*{0.2in} Abhik Roychoudhury\\
National University of Singapore\\
{\tt \{hruan,yuntong,abhik\}@comp.nus.edu.sg}
}



\maketitle
\thispagestyle{plain}

\begin{abstract}
Autonomous program improvement typically involves automatically producing bug fixes and feature additions. Such program improvement can be accomplished by a combination of large language model (LLM) and program analysis capabilities, in the form of an LLM agent. Since program repair or program improvement typically requires a specification of intended behavior - specification inference can be useful for producing high quality program patches. In this work, we examine efficient and low-cost workflows for iterative specification inference within an LLM agent. Given a GitHub issue to be resolved in a software project, our goal is to conduct iterative code search accompanied by specification inference - thereby inferring intent from both the project structure and behavior. The intent thus captured is examined by a reviewer agent with the goal of vetting the patches as well as providing a measure of confidence in the vetted patches. Our approach \toolname is built on the open-source LLM agent AutoCodeRover. In an evaluation on the full SWE-Bench consisting of 2294 GitHub issues, it shows more than 50\% improvement in efficacy over AutoCodeRover. 
Compared to the open-source agents available, our work shows modest cost (\$0.65 per issue) in resolving an average GitHub issue in SWE-Bench lite. The production of explanation by \toolname allows for a better ``signal" to be given to the developer, on when the suggested patches can be accepted with confidence. 
\toolname also seeks to demonstrate the continued importance of specification inference in automated program repair, even as program repair technologies enter the LLM era.
\end{abstract}


\section{Introduction}

Automatic programming has long been an aspiration of software engineering research. It has inspired research in topics like program synthesis and repair. In recent times, automatic programming from natural language specifications has become somewhat more realistic due to the emergence of tools like GitHub Copilot. At the same time, the automatically generated code from Large Language Models (LLMs) suffers from errors and vulnerabilities \cite{Fan23,Karri22} and needs to be improved. 
For this reason, there has been a recent research focus on autonomous program improvement. The problem setting for autonomous program improvement
involves solving of GitHub issues which would typically involve bug
fixes or feature additions.
Though these tools are employed on manually written software projects such as the recently proposed SWE-bench \cite{jimenez2024swebench}, they hold the promise of high quality trustworthy code construction from LLMs. Starting with the AI software engineer Devin \cite{devin} from a stealth startup, recently several autonomous program improvement tools such as AutoCodeRover \cite{autocoderover} have been proposed for automatically solving GitHub issues (such as bug fixes or feature additions). By combining these technologies with code generation via GitHub Copilot, one can envision trustworthy code construction from LLMs. 

Program improvement or program repair, typically requires capturing developer {\em intent} to guide the process.  However, there is no formal specification of developer intent. The natural-language description of the developer intent is usually only available at a 
``higher level" - it captures the intended behavior of the entire software system. However to improve or repair specific components of a software system (where the error might have been localized) - one needs to infer specifications of the different components. A successful approach to program repair may thus involve
specification inference - where by carefully analyzing the 
artifacts of the buggy program (such as program executions), 
we can infer snippets of the intended
program behavior.  The works
on semantic program repair \cite{semfix,angelix} extract specifications via symbolic analysis of the given tests. Indeed, the existing literature on program repair \cite{cacm19} uses a given test-suite as developer intent, and hence is focused on avoiding test-data over-fitting. The works on semantic repair alleviate the over-fitting concern by inferring symbolic specifications from tests. 
Nevertheless, for the general problem of program improvement, the buggy program may or may not be accompanied by tests. Moreover, symbolic analysis based program repair has a high entry barrier for developers. For these reasons, recently autonomous program improvement using Large Language Models (LLMs) \cite{autocoderover,swe-agent,agentless}
has been studied.

In this work, we explore the role of program specifications thoroughly in LLM-guided autonomous software engineering workflows. 
To understand the intent of the developer and perform program improvement based on inferred specifications, we build our work on the publicly available AutoCodeRover \cite{autocoderover} tool. The reason for this choice is strategic. In
essence, AutoCodeRover takes the position that the structure of the
program also captures a coarse encoding of the developer intent, and it
tries to glean intent by analyzing (and searching over) the program
structure; it performs code search on the project structure for fix localization. 
Thus, to build a workflow where we conduct high quality program improvement via iterative specification inference, we choose to build our work on AutoCodeRover. 
Our work looks into various sources of specifications such as function-level code summaries and testcases, apart from program structure. 
The core contribution thus lies in distilling the various specifications coming from different sources into a single patch.

We thus present \toolname, a progeny of AutoCodeRover, which conducts and exploits more powerful specification inference.
Starting from a GitHub issue, it conducts code search guided by the
program structure, as in AutoCodeRover. However, in the process of the
code search, as it visits classes/methods, it also calculates and
deposits the specifications of the classes/methods which would have
allowed for remediation of the observable error, thereby capturing {\em intended} program behavior. The specifications gathered from the code
search are deposited along with generated tests to a reviewer agent. The
reviewer agent studies the specifications, generated tests, and natural language requirements to guide the patching. More importantly, the
reviewer agent produces evidence of confidence in the reported patch.

\paragraph*{Contributions}
The core contributions of our work on \toolname can be summarized as follows. 
\begin{itemize}
\item {\em Specification Inference:} We examine the role of specification inference in LLM guided autonomous software engineering. Our work suggests iterative specification inference to guide patching in LLM oriented program repair workflows. 
Once the understanding of developer intent is accomplished via iterative specification inference, patch construction is a natural by-product of the inferred specification.


\item {\em Suggesting patches with confidence: } We design a reviewer agent for code review which reconciles specifications, tests and natural language requirements.  
The reviewer agent can be seen as conducting a comprehensive patch validation. The reviewer agent can  produce evidence of correctness of automatically generated fixes -  such as explanation of patch, reproducer test, and the accumulated specifications from different code elements. These evidence can be maintained along with the automatically generated patches, to track future regressions. 

\item {\em Experimental evidence:} Our tool shows high efficacy, solving $19.3\%$ issues in full SWE-bench and $31\%$ on SWE-bench lite. We also balance other needs from LLM agents such as low cost (\$0.65 per issue) and supporting higher precision/recall. Our tool is available open-source in Zenodo and will be publicly released upon acceptance.
\end{itemize}




\section{Motivating Example}

\begin{figure}[ht]
    \centering
    \begin{subfigure}[b]{\columnwidth}
        \centering
\includegraphics[trim=2.5mm 236.5mm 91.5mm 1.5mm, clip,width=0.8\columnwidth]{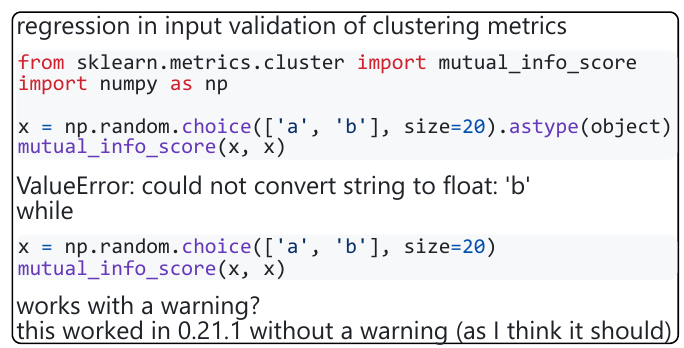}
    \vspace{-5pt}
    \caption{Issue statement.}
    \label{fig:scikit-15535}
    \end{subfigure}
    
    \vspace{5pt}
    
    \begin{subfigure}[b]{\columnwidth}
        \centering
\includegraphics[trim=4mm 251mm 88mm 3mm, clip,width=0.8\columnwidth]{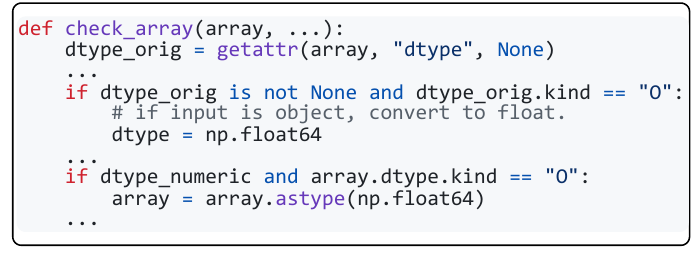}
    \vspace{-5pt}
    \caption{Buggy function.}
    \label{fig:check-array}
    \end{subfigure}


    \begin{subfigure}[b]{\columnwidth}
        \centering
        \includegraphics[trim=0in 10.69in 3.82in 0in, clip,width=0.8\columnwidth]{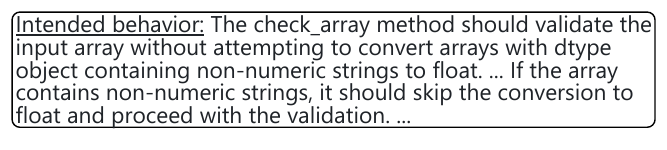}
        \vspace{-10pt}
        \caption{Inferred intended behavior of the  buggy function.}
        \label{fig:scikit-15535-func-summary}
    \end{subfigure}

    \vspace{5pt}

    \begin{subfigure}[b]{\columnwidth}
        \centering
        \includegraphics[trim=0mm 261.5mm 98.5mm 3mm, clip,width=0.8\columnwidth]{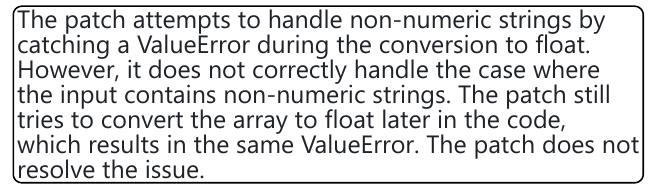}
        \vspace{-5pt}
        \caption{Reviewer feedback on initial patch.}
        \label{fig:scikit-15535-feedback}
    \end{subfigure}

    \vspace{5pt}
    
    \begin{subfigure}[b]{\columnwidth}
        \centering
        \includegraphics[trim=4mm 196mm 60mm 4mm, clip,width=0.8\columnwidth]{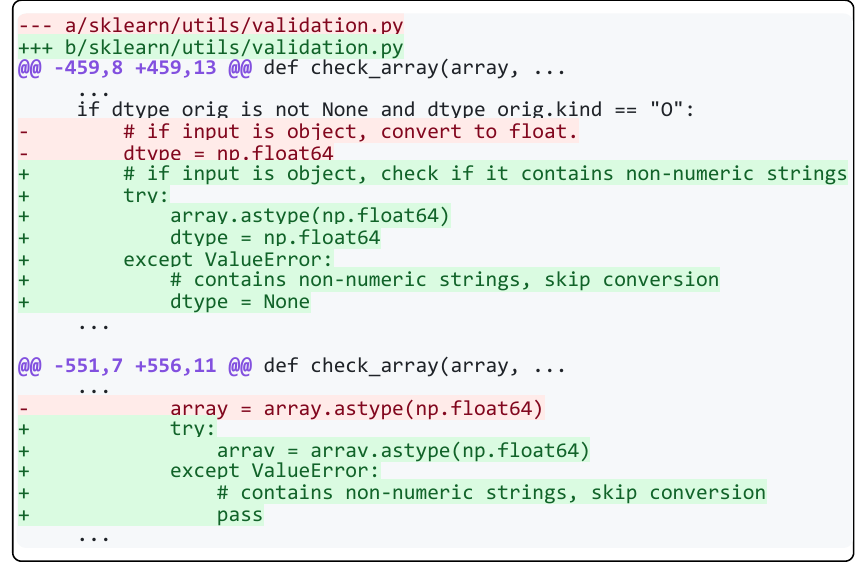}
        \vspace{-5pt}
        \caption{Final patch generated by \toolname. 
    }
        \label{fig:scikit-15535-patch}
    \end{subfigure}

\caption{scikit-learn-15535 description and \toolname artifacts.}


\end{figure}





We now present the \toolname approach via an example. 
The GitHub issue involved in this example is scikit-learn-15535\footnote{\url{https://github.com/scikit-learn/scikit-learn/issues/15534}}, shown in Figure~\ref{fig:scikit-15535}. In the issue, two relevant code snippets are provided. According to the issue report, both snippets had worked without problem on an older version of scikit-learn, and it is expected that they continue to work on the current version. However, on the current version, the first snippet now crashes. The associated error information indicates that the crash occurred when scikit-learn mistakenly tries to convert a non-numeric array element into a float.

To resolve the issue, \toolname first identifies buggy program locations by exploring the program and retrieving relevant code. In this example, the 
identified buggy method is \code{check\_array}, shown in Figure~\ref{fig:check-array}. The method performs two conversions of string to float. The two conversions are the root cause of the failure reported in the issue, resulting in an exception when the involved string is non-numeric. 
\toolname inferred a summary of \textit{intended behavior} of the method \code{check\_array} (shown in Figure~\ref{fig:scikit-15535-func-summary}), which serves as a specification of how the method should be modified.

Next, the identified buggy method and its intended behavior are passed to our patching agent, which will write patches for the method. At the first attempt, the patching agent wrote a partial patch, which only contains the first of the two hunks in Figure~\ref{fig:scikit-15535-patch}, i.e., the patch only catches exceptions for the first conversion. Without further rectification of the patch, the issue could not be resolved.
To vet the patch for potential mistakes, the initial patch is then passed to the reviewer agent. Apart from the patch, the reviewer agent also takes a reproducer test that reproduces the issue.
The reviewer agent then runs the reproducer test on both the original program and the program repaired by the initial patch. By referring to the error information, the patch, and the issue statement, the reviewer agent is able to give the feedback as shown in Figure~\ref{fig:scikit-15535-feedback}. The feedback correctly indicates that the initial patch does not resolve the issue and can be rectified by catching exceptions for the other string conversion. Finally, the feedback is passed to the patching agent, which writes the correct patch shown in Figure~\ref{fig:scikit-15535-patch}.

In this example, we illustrated how our reviewer agent provides feedback on an incorrect patch for our patch-writing agent. The feedback leads to a later rectification of the patch, and explains clearly why the initial patch is incorrect.


\section{Methodology}

\begin{figure*}[t]
    \centering
    \includegraphics[trim=0.3cm 0 0.2cm 0, clip, width=0.88\textwidth]{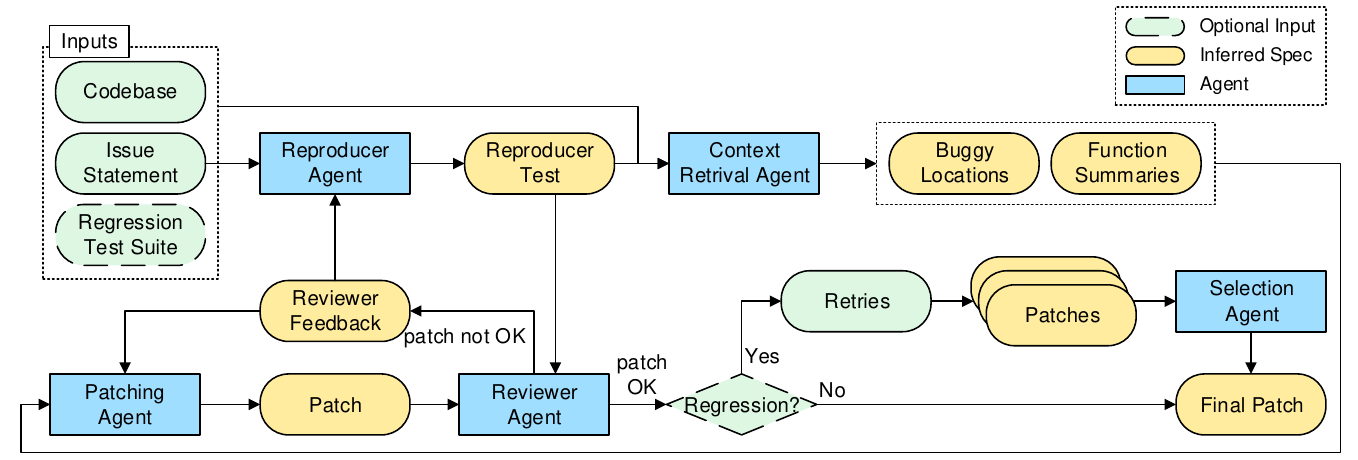}
    
    \vspace{1em}
    
    \caption{Overall Workflow of \toolname.}
    \label{fig:workflow}
\end{figure*}

\subsection{Overview}

\paragraph*{Problem setup} Given a software codebase $C$ and a natural language problem description $D$, the goal is to automatically derive a patch $p$ (i.e. a set of code modifications) to $C$, such that the patched codebase $C'$ satisfies the requirements in $D$.
One example setup for $D$ is GitHub issues, in which the issue description contains requirements for fixing a bug or adding a new feature.

In this paper, we drive autonomous program improvement with the help of program specifications. We try to acquire an understanding of  the intended program behavior (the specification), which then allows us to produce high-quality patches that successfully resolve GitHub issues. Beyond producing high-quality patches, an additional benefit of understanding the specification is that it also serves as evidence as to \emph{why} the patch is correct. The evidence holds promise in terms of easing software maintenance and engendering trust in the code. 
The key novelty of our approach lies in how we infer and utilize various forms of specifications. For an overview of all the specifications involved, we depict the general workflow of our approach \toolname in Figure~\ref{fig:workflow}. In this figure, the inferred specifications are highlighted in yellow. We also highlight in blue all the LLM agents present in the workflow. As shown in Figure~\ref{fig:workflow}, the specifications are inferred in an iterative fashion: the agents take in specifications, possibly produced by other agents, and in turn infer new forms of specifications. This iterative process generates a variety of specifications, until a patch is generated and deemed correct by one of our agents that vets generated patches. 

Specifically, as shown in Figure~\ref{fig:workflow}, the following specifications are inferred in sequence in \toolname, which is given as input an issue statement and a software codebase.
\begin{enumerate}
    \item The input issue statement is passed to a reproducer agent, which writes a \emph{reproducer test} that reproduces the program fault reported in the issue.
    \item The reproducer test, its execution results, along with the issue statement and the codebase, are passed to a context retrieval agent. 
    The context retrieval agent explores the program codebase and identifies the relevant code to the issue. 
    It eventually decides on a set of \emph{buggy locations} that need patching.
    \item \label{spec:summary} The context retrieval agent also produces a \emph{function summary} for every function encountered while exploring the program code. A function summary describes the intended behavior of a function in natural language, with respect to the current issue being solved.
    \item The buggy locations, together with their corresponding function summaries, are passed to a patching agent, which tries to write a \emph{patch} to resolve the issue.
    \item \label{spec:reviewer} The patch and the reproducer test are passed to a reviewer agent for scrutiny. The reviewer agent will produce a \emph{reviewer feedback} if the patch is deemed incorrect; the patching agent will take in the reviewer feedback and try writing another patch. The reviewer feedback is a natural-language explanation of why the patch is incorrect and how it can be rectified. Likewise, a reviewer feedback for the reproducer test will be produced at the same time if the test is deemed incorrect.
    \item If a patch is deemed correct by the reviewer agent, and there is an existing regression test suite available for the program, the patch will be checked via the regression test suite. If there is no regression, the patch will be accepted as the final patch.
    Otherwise, if some of the regression tests fail, we will retry the workflow up to a predefined number of times.
    \item \label{spec:selection} Finally, after multiple retries, there can be multiple patch candidates. A selection agent is invoked to select one final patch among the patch candidates, and give the reason why this patch is selected. The final patch, the reason for selection, and optionally the rest of the candidate patches will be sent to the user.
\end{enumerate}


Among the specifications, the \ref{spec:summary}) function summary and  \ref{spec:reviewer}) reviewer feedback are unique to \toolname and unexplored by other LLM agents. These specifications have boosted the effectiveness of \toolname in resolving software issues, because they fully exploit different kinds of software artifacts: the function summary exploits the program code behavior, and the reviewer feedback exploits both the code and the test. 

\subsection{Function Summary: Specification from Program}
\label{sec:summary}

\begin{figure}[t]
    \centering
    \includegraphics[ width=\columnwidth]{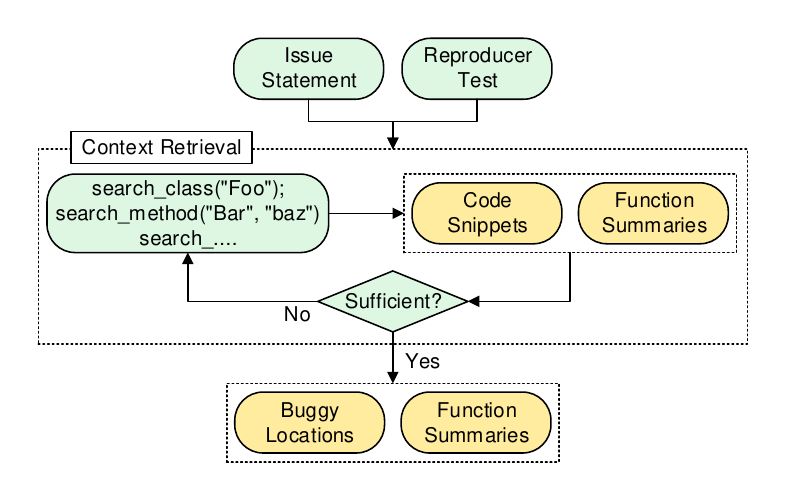}
    \caption{Context retrieval in \toolname.}
    \label{fig:retrieval}
\end{figure}

In this section, we first describe how the context retrieval agent gathers code context for the software issue to be resolved. We then discuss how \toolname transforms the user intent in the issue description into program specifications for shorter code elements such as functions.

Existing LLM programming agents typically employ a \textit{context retrieval} step to collect necessary code context related to the given issue from a large codebase.
\toolname follows the general architectural design of programming agents in its context retrieval stage, as shown in Figure~\ref{fig:retrieval}.
\toolname conducts context retrieval by providing a set of APIs to the LLM for exploring the codebase.
The LLM agent invokes the retrieval APIs to investigate the relevant code snippets in the program. 
The retrieved code forms the code context for the current to-be-resolved issue, which can contain definitions of the relevant classes and methods.
After each round of retrieval API invocations, the LLM agent takes the code context collected so far and decides whether the context is sufficient for understanding and resolving the problem.
If the context is deemed sufficient, the retrieval process will end, and the agent will decide on a set of \textit{buggy locations}, which are sent to the patching agent for repairing. Otherwise, the retrieval process continues until a predefined threshold count is reached.

One key novelty in \toolname is the \textit{explicit} extraction of \textit{function summaries} while collecting code snippets during context retrieval.
In \toolname, whenever a new code snippet is retrieved with an API and sent to the context retrieval agent, we explicitly prompt the agent to analyze the ``intended behavior'' of this code snippet in the current problem context.
The intended behavior (or specification) is a concise natural-language summary of how a function should behave to meet the requirements specified in the high-level problem description.
This function-level summary of intended behavior serves as a local specification to guide the patch construction. 
The system-level intended behavior specification given by the user (i.e. the issue description) is often on how the program should behave rather than how a unit function should behave. So we usually do not have the intended behavior of a function.
Although the issue description may provide some ``direction'' on the intended behavior of a function - it is usually not sufficient to guide the patching agent.
On the other hand, the extracted function-level specification (capturing the intended behavior of the function) serves as a more direct guide to the patching agent. 
Instead of giving a set of bug locations $\{L_{1}, L_{2}, ..., L_{n}\}$ to the patching agent to modify, \toolname gives the pairs of bug locations and their corresponding local specification $\{ (L_{1}, Spec_{1}), (L_{2}, Spec_{2}), ..., (L_{n}, Spec_{n}) \}$.
The patching agent can then refer to the  specifications of intended behavior and modify code at the function level (so as to achieve this intended behavior).
Intuitively, our approach decomposes the repository-level issue solving task to several function-level code modification tasks, in which each function-level task has a natural language specification.
LLMs have been extensively studied for function-level coding tasks and have shown promising results in function-level benchmarks such as HumanEval~\cite{humaneval,evalplus} and MBPP~\cite{mbpp}.
Therefore, this task decomposition helps the patching agent of \toolname
which then has to solve  smaller and more manageable tasks.

\subsection{Reviewer Feedback: Reconciling Specifications}
\label{sec:reviewer}

Another kind of specification inferred by \toolname is the reviewer feedback. 
To be more precise, the reviewer feedback can be called a meta-specification: it is a reflection on the specifications inferred in previous steps. Concretely, given a patch and a reproducer test, the reviewer agent in \toolname will provide feedback, which includes 1) a binary decision of whether the patch and the reproducer test are correct respectively; and 2) an explanation for the decisions.

The reviewer feedback contributes to our specification inference practice in two ways. First, it makes the specification inference iterative. The reviewer feedback will be passed back to the patching agent and the reproducer agent, leading to improved patches and reproducer tests.
Second, it reconciles the patch and the test. In this way, errors that are not obvious when examining the two separately can be revealed and rectified.
What makes the reviewer feedback important is the absence of a suitable test suite. If a test suite was available for checking whether the issue has been resolved, patch correctness could be easily decided. In reality, however, issues occur when the program already passes the accompanying regression tests, which means that a high quality test-suite to check a generated patch (for the given issue) is usually not available.

To mitigate the lack of an issue-revealing test case, \toolname writes a reproducer test via the reproducer agent. However, this test alone is not sufficient for deciding patch correctness. This is because the reproducer test can be \emph{incorrect}, due to the non-determinism of the LLM, i.e., the test may fail a patch that actually conforms to the user intent.
Besides, the reproducer test can also be \emph{incomplete} description of intent, i.e., a patch may pass the test without completely resolving the issue.
The limitation of the reproducer test derives from the fact that tests are a precise yet incomplete specification. To overcome the limitation, we make the observation that the natural-language issue statement is ambiguous in nature yet often contains richer information. Therefore, supplementing the test with an understanding of the issue statement is likely to help decide patch correctness. This is accomplished in the reviewer agent of \toolname, which considers the issue statement as well as test to vet patch candidates.

Beyond deciding on patch correctness, the more important aspect of the reviewer feedback is an explanation of the decision made, which will help the patch agent rectify an incorrect patch. Further, for a correct patch, the explanation will help the user understand and accept the patch. The user can merge the reviewer feedback into the software together with the patch for future reference, which will help software maintenance in the long run.

\begin{figure}[ht]
    \centering
    \includegraphics[trim=0mm 239.5mm 43.5mm 4mm, clip,width=0.9\columnwidth]{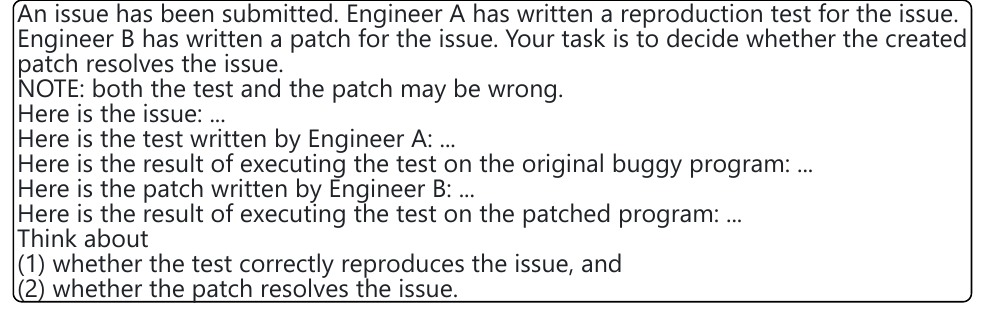}
    \caption{Template prompt for the reviewer agent.}
    \label{fig:reviewer-prompt}
\end{figure}

Our reviewer agent generates the feedback in two steps. First, the original program and the patched program are run on the reproducer test. In both runs, execution information including the output and the exit code are collected. The reviewer agent then provides the LLM with these execution information, along with the issue statement and the reproducer test. The LLM is prompted to decide whether the patch and the test are correct respectively, and to provide explanations for both decisions. Figure~\ref{fig:reviewer-prompt} shows the template of the prompt.

\subsection{Patch Selection}

As shown in Figure~\ref{fig:workflow}, a patch approved by the reviewer agent is checked through a regression test suite, which serves as an oracle for whether the patch breaks existing functionality of the program.
However, in the setting of resolving GitHub issues, the regression test suite can be an inaccurate oracle, meaning that they can reject correct patches which resolve the issue.
This is because the correct patch may inevitably modify existing functionalities of the program while resolving the issue, thus causing some of the existing regression tests to fail.
For example, if the patch needs to modify the signature of an existing function \texttt{f} in order to resolve an issue, regression tests that invoke \texttt{f} will now fail.
Since the correct patch can be rejected by the regression tests, 
we employ a patch \textit{selection} process at the end of the workflow to select the most promising patches among the rejected candidate patches.

During the final patch selection phase, \toolname goes beyond the test cases and employs a \textit{selection agent} to choose a patch based on the natural language issue description. 
All candidate patches that failed some tests are presented to the selection agent, together with the issue description.
The selection agent is instructed to analyze the root cause of the issue, think about how the issue can be possibly resolved, and select a patch that best addresses the issue.
This natural language-guided patch selection can recover correct patches that are mistakenly filtered out by an inaccurate test suite. 
It exploits the natural language issue report as that captures the most up-to-date intents from users/developers.
While making a choice among the candidate patches, the selection agent also explicitly states a reason why it chooses a particular patch among the candidate patches. 
This ``reason for selection'' can be given as {\em evidence} together with the final patch.

\subsection{Evidence}
\label{sec:evidence}

\toolname is designed to not only generate a patch to resolve the issues in software repositories, but also to provide the inferred specifications as evidence for why a patch was selected.
Specifically, along with the final patch, the following artifacts can be the outputs of \toolname as well:

\begin{itemize}
    \item Buggy locations and their intended behaviors.
    \item The reproducer test written by the reproducer agent.
    \item The reason why the final patch was approved by the reviewer agent (if the patch is approved by the reviewer and the regression test suite).
    \item The reason why the final patch is selected by the selection agent (if there are multiple candidate patches which cannot be differentiated by the tests).
\end{itemize}

The benefits of generating evidence are threefold.
First, these artifacts can guide the LLM agents in constructing higher quality patches, as discussed in Section~\ref{sec:summary} and \ref{sec:reviewer}.
Second, the natural language artifacts can assist the developers in understanding the auto-generated patches more quickly. 
Before examining the actual patch, developers can gain a preliminary understanding of the changes by reviewing the reasons for approval or selection and the intended behaviors for the buggy locations.
Last but not least, the evidence can be integrated into the software repository and can evolve with it.
The developers can integrate the reproducer test for this issue as part of the test-suite of the program.
Reasons for patch approval/selection can become parts of the commit message when the auto-generated patch is committed to the repository.
Overall, we propose \toolname as a programming agent that not only automatically generates code improvements but also produces evidence that enriches the software system lifecycle.




\section{Experimental Setup}

We address the following research questions:
\begin{description}[leftmargin=*]\itemsep0em 
    \item[RQ1:] What is the efficacy of \toolname in resolving  issues?

    \item[RQ2:] What level of confidence can developers get from the patch and specifications produced by \toolname?

    \item[RQ3:] What is the quality of the specification produced by SpecRover as evidence? 




\end{description}

\paragraph{Benchmark} 
We evaluate the efficacy of \toolname on SWE-bench~\cite{jimenez2024swebench}, a widely-used benchmark for autonomous program improvement consisting of 2294 real-world GitHub issues. 
For each issue, the only input for \toolname is the issue statement and the 
buggy codebase. 
Note that the regression test suite used by \toolname is part of the \emph{buggy} program; we do not access any code or test that is added by the developer in the fixed version of program. 

\paragraph{Baselines and Evaluation Metrics} 
For RQ1, we compare with the state-of-the-art systems that target the repository-level issue solving task. 
In our comparison, we include all the open-source software engineering agents which have reported results on \swebench. The baseline tools include:


\begin{itemize}[leftmargin=*]
    \item AutoCodeRover \cite{autocoderover}. AutoCodeRover employs a set of program structure-aware APIs to gather relevant code context. It optionally integrates debugging techniques such as Spectrum-based Fault Localization to sharpen the context.
    
    \item SWE-agent \cite{swe-agent}. SWE-agent designs an agent-computer interface, which defines the possible actions taken by an agent to edit code, navigate the codebase, and execute tests.

    \item AppMap Navie \cite{appmap}. Navie uses a retrieval-augmented generation (RAG) based approach to construct the code context, and performs an explicit planning step before generating code changes~\cite{appmap-report}.

    \item OpenDevin \cite{opendevin}. OpenDevin's CodeActAgent tackles the tasks by having a general action space, where the agent is allowed to execute arbitrary Python and bash commands inside a sandbox environment. 

    \item Aider \cite{aider}. Aider constructs a repository map which helps the LLM to understand the repository context. It also uses the regression test suite as a harness to retry the task.

    \item Moatless Tools \cite{moatless}. Moatless Tools builds an agentic loop that functions as a finite state machine and transitions between states. It focuses on building a set of good tools for the agent instead of relying on the agent for reasoning.
    
    \item Agentless \cite{agentless}. Agentless is a concurrently pursued (currently unpublished) arXiv report which employs a fixed two-phase approach of localization and repair, without allowing the LLM to decide on actions or utilize tools.

\end{itemize}

We report pass@1 efficacy on \swebench for all tools.
For each issue, \swebench has a set of acceptance tests written by the developers to evaluate the patch correctness.
These acceptance tests are not used by the tools when generating patches.
We follow the official \swebench evaluation criteria - if the single patch generated by a tool passes the \swebench acceptance tests for the issue, the issue is considered as resolved.


\paragraph{Implementation and Parameters}

We implemented \toolname on top of the AutoCodeRover codebase and reuse its context retrieval APIs.
We implemented new features unique to \toolname such as function summary extraction as part of the context retrieval process.
Other unique features such as patch reviewing and selection are implemented as new LLM agents.
\toolname supports multiple LLMs as backend. 
In our experiments, we used the Claude 3.5 Sonnet as the main foundation model, and only switch to OpenAI GPT-4o for a task if that task encounters an API error when invoking the Claude remote APIs.
We set maximum retries on regression test suite failures to be $3$.

\paragraph{Randomness of LLMs}
LLMs are inherently random in its output generation, which may threaten the validity of LLM-based coding agents including \toolname. 
We address this by setting the model temperature to 0, so that the model output is more deterministic.

\paragraph{Manual Inspection of Results}

\new{For a better understanding of the experimental results, we perform manual inspection to obtain certain data, e.g., patch overfitting rate. All manual inspections were independently conducted by two authors and subsequently cross-checked. Discrepancies were resolved through consultation with a third colleague from our group.}



\section{evaluation}

\begin{table*}[t]
\centering
\begin{minipage}{0.7\linewidth}
\centering
\footnotesize 
\caption{Comparison of 
efficacy/efficiency/cost on \swebench and \swebenchlite.}
\label{tab:efficacy}

\begin{tabularx}{\textwidth}{l r | r | X | X}
\toprule
Tool       & LLM            & Resolved\%        & Avg. Time (s) & Avg. Cost (\$) \\
\midrule
\midrule
\multicolumn{5}{c}{Efficacy on \swebench (size=2294)}\\

\midrule

AutoCodeRover & GPT-4   &  12.42\% (285)     & 248   & 0.45 \\
SWE-Agent     & GPT-4   &   12.47\% (286)    &  -  & 1.59   \\
AppMap Navie  & GPT4o   &  14.60\% (335)     &  -  & -      \\
\rowcolor{blue!15} \toolname     & Sonnet-3.5+GPT-4o          & \textbf{19.31\%} (443) &  362  &   0.72     \\

\midrule
\midrule 

\multicolumn{5}{c}{Efficacy on \swebenchlite (size=300)}\\

\midrule

SWE-Agent     & GPT-4           & 18.00\% (54)      &   -   & 1.67 \\
AutoCodeRover & GPT-4           & 19.00\% (57)      &  195  & 0.43 \\
AppMap Navie  & GPT-4o          & 21.67\% (65)      &   -   &  -   \\
OpenDevin     & Sonnet-3.5      & 26.00\% (78)      &  -    & 1.10 \\
Aider         & GPT-4o+Opus-3   & 26.33\% (79)      &   -   &  -   \\
Moatless Tools    & Sonnet-3.5      & 26.67\% (80)      &  71  & 0.17 \\
Agentless     & GPT-4o          & 27.33\% (82)      &   -   & 0.34 \\

\rowcolor{blue!15} \toolname     & Sonnet-3.5+GPT-4o      & \textbf{31.00\%} (93) & 309  & 0.65 \\
\bottomrule
\end{tabularx}
\vspace{-5pt}
\footnotetext{`-' indicates data is not publicly available.}
\end{minipage}
\end{table*}

\subsection{RQ1: Overall Efficacy of Task Resolving}

We first evaluate the overall efficacy of \toolname in resolving repository-level tasks.
We report the efficacy of \toolname on both \swebench (consisting of 2294 real-world GitHub issues), and \swebenchlite (which is a subset of \swebench consisting of 300 issues).
For the baseline tools, we compare with their corresponding reported efficacy.
If a tool supports different configurations (e.g. different LLMs as the backend), we compare with the configuration with the highest efficacy.

\begin{figure}[t]
    \centering
    \includegraphics[width=0.75\columnwidth]{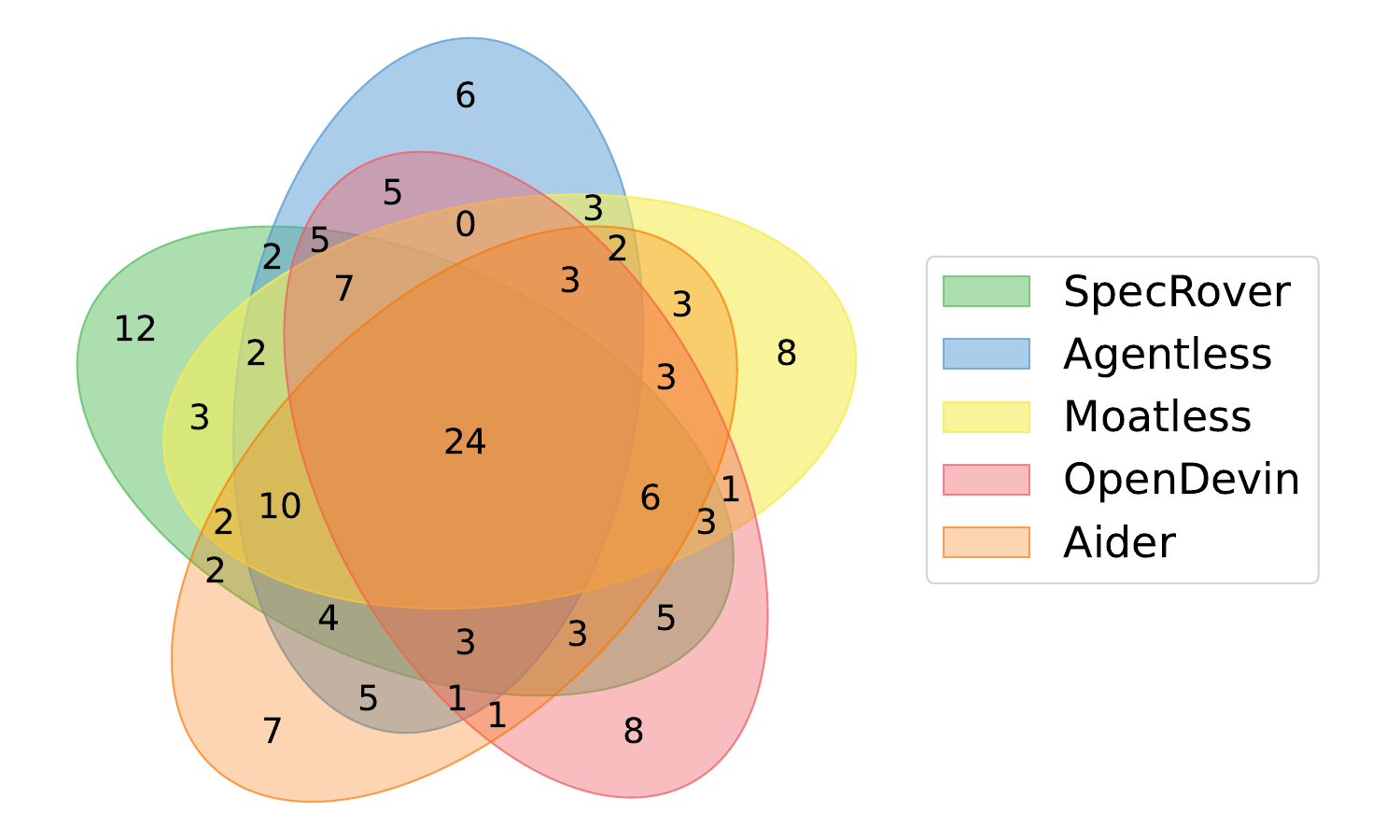}
    \caption{Number of uniquely resolved issues by the top performing open-source tools on \swebenchlite.}
    \label{fig:venn-lite}
\end{figure}

\paragraph*{Result}
Table~\ref{tab:efficacy} shows the efficacy of issue resolving in both \swebench and \swebenchlite.
Overall, \toolname achieves the highest efficacy among all the open-source tools in both \swebench and \swebenchlite. 
In \swebenchlite, compared to the previously top-performing group of tools 
which resolved approximately 26\% to 27\% of the issues, \toolname improved the efficacy to 31\%.
This efficacy improvement is also evident in the full \swebench, where \toolname improved the efficacy from 14.60\% to 19.31\%.  
Figure~\ref{fig:venn-lite} illustrates the number of uniquely resolved issues by \toolname and other tools in \swebenchlite.
For clarity, this figure includes only the top five performing tools from Table~\ref{tab:efficacy}.
\toolname uniquely resolved 12 issues, the highest number of uniquely resolved issues among all the tools.
Among the 12 uniquely resolved issues, \toolname resolved six of them by generating only one patch, demonstrating that the inferred function summary can effectively guide the LLM to generate correct patches.
For the other six issues, \toolname deemed the first generated patch as incorrect from the reviewer agent and the regression test suite.
In this case, the patches are iteratively refined based on the reviewer feedback and the test results, and eventually the correct patch is selected at the end of the workflow.



\paragraph*{\new{Data Memorization}}
\new{
An issue that can bias the evaluation of LLM agent-generated patches is \emph{data memorization}~\cite{Magar022contamination}. Data memorization  occurs when an LLM deals with a program that exists in its training set. To evaluate the risk of data leakage, we specifically count the patches generated by \toolname that are syntactically identical to the ground-truth patches. Our counting shows that in SWE-Bench Lite, \toolname generated the ground-truth patch only for 9 issues, accounting for 10\% of the 93 resolved issues. The result shows that data memorization occurs very infrequently for \toolname.
}

\paragraph*{Time and Cost}

We also report the average time taken and average costs for each issue in Table~\ref{tab:efficacy}.
For each tool, we include the time and cost statistics in Table~\ref{tab:efficacy} if these information was publicly reported or can be calculated from their publicly available execution traces.
On average, \toolname costs \$0.65 USD to generate patches for each issue in \swebenchlite, achieving the highest efficacy with a relatively low cost.
We further investigate the 93/300 issues resolved by \toolname in SWE-bench lite.
For the resolved issues, \toolname only costs \$0.36 USD per issue to generate the correct patch, suggesting that the resolved issues requires less retries and less API calls to the LLM in general.
Time-wise, \toolname spends an average of 309 seconds (i.e. 5.15 minutes) on each issue, which includes the time for executing the reproducer and the regressions tests in the project.
According to a recent study, most developers accept automated program repair tools which takes less than 30 minutes~\cite{noller2022trust}.
\toolname requires approximately 5 minutes, which we deem acceptable.

\paragraph*{\new{Patch Correctness}}
\new{
A patch that passes a given test suite is not necessarily correct, because the test suite is an incomplete specification. This problem is known as \emph{patch overfitting}~\cite{edward15overfitting} in program repair. For a more accurate evaluation of the generated patches, we manually compared the \toolname-generated patch with the developer patch for each resolved issue. Our manual investigation confirmed that 60.2\% (56/93) of the patches that pass the test suites are semantically equivalent to the ground-truth patches. We also observed that 29 out of the 37 (=93-56) overfitting patches (i.e., the patches that pass the test suite but are not semantically equivalent to ground truth) modify the same methods as the ground-truth patch, and only the specific modification is semantically different from the ground truth. This implies that even overfitting patches produced by \toolname are useful in indicating a fix location. 
Adding up the semantically equivalent patches and the overfitting patches that have the correct fix location, a total of over 90\% of the test-passing patches prove useful for issue resolution.

\paragraph*{\new{Reasons for Failure}}
\new{
We examined the 207 tasks that are not resolved by \toolname in SWE-Bench Lite.These failure cases break down into three cases:

\begin{enumerate}
    \item Ambiguous issue description (as labeled by SWE-bench Verified~\cite{chowdhury2024verified}): 107 (51.7\%)

    \item Incorrect fix location: 61 (29.5\%)

    \item Incorrect code modification: 39 (18.8\%)

\end{enumerate}

Case~1 is ``hard or impossible to solve''~\cite{chowdhury2024verified} without further information. Case~2 is more frequent than Case~3, implying that \toolname can potentially have significant improvement by leveraging more fix localization techniques. For Case~3, in most such cases (26/39), we observed that the gist of the patch generated by \toolname is actually correct, but some code detail is wrong. To reduce this kind of failures, more test generation techniques can be leveraged within our agent.

}
}

\subsection{RQ2: Utility of autonomous SE, confidence in results}

Although the efficacy in resolving issues is an important aspect of autonomous program improvement, it is not the sole purpose of such a technique. Rather, the efficacy is a means to an end -- to reduce human effort in software maintenance. To this end, a program improvement technique must not only have high efficacy, but also minimize the effort required of an end user to use the technique. The effort is related to two metrics: 1) \emph{signal-to-noise ratio}, i.e., the ratio of correct to incorrect patches presented to a user; and 2) the difficulty of examining each auto-generated patch that is suggested.

We have designed \toolname to reduce both of these efforts. First, to reduce the number of incorrect patches that a user may examine, we use the reviewer agent to decide the correctness of the generated patch and the reproducer test. The user can choose to examine the generated patch only when both the patch and the reproducer test are deemed correct by the reviewer agent. The accuracy of the reviewer decisions are measured in RQ2. Second, to make it easy for a user to examine each patch, \toolname provides a variety of evidence to help understand the patch, as discussed in Section~\ref{sec:evidence}. The quality of the evidence will be discussed in RQ3.

There can be four different scenarios when the reviewer decision is viewed in relation to the actual correctness of the patch. For convenience, we say a patch is \emph{accepted} when the reviewer agent decides that both the generated patch and the reproducer test are correct. With this, we discuss the following four scenarios:

\begin{table}[h!]
\centering
\footnotesize
\caption{Reviewer decisions on SWE-Bench lite.}
\label{tab:precision}
\begin{tabular}{ll}
\toprule
Category & \# Tasks \\
\midrule
TP & 26 \\ 
TN & 51  \\
FP & 26 \\ 
FN & 16 \\
\midrule
Total & 119 \\
\midrule
Accuracy\,=\,(TP+TN) / Total & 64.7\% \\
Precision\,=\,TP / (TP+FP) & 50.0\% \\
Recall\,=\,TP / (TP+FN) & 61.9\% \\
\bottomrule
\end{tabular}
\end{table}

\begin{itemize}
    \item TP (true positive): Patch is accepted and correct;
    \item TN (true negative): Patch is not accepted and incorrect;
    \item FP (false positive): Patch is accepted but incorrect;
    \item FN (false negative): Patch is not accepted but correct.
\end{itemize}

Table~\ref{tab:precision} lists the frequency that each scenario occurred in our experiment. The table counts in 119 tasks in \swebenchlite for which a reproducer test was generated. In the table, we also calculate the accuracy, precision, and recall of the reviewer decisions. Out of the 119 tasks, there are 26 TP's and 51 TN's, i.e., as many as 64.7\% (accuracy) of the reviewer decisions were consistent with the actual correctness of the patch. The recall was also over 60\%, meaning that the majority of the generated correct patches were recognized by the reviewer agent.

\begin{figure}[t]
    \centering
    \includegraphics[width=0.95\columnwidth]{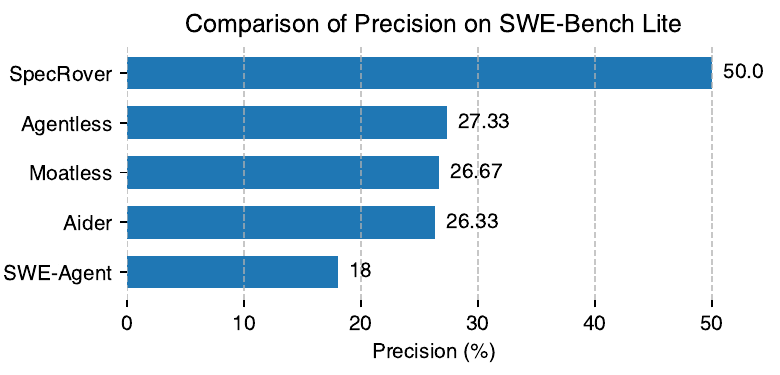}
    \caption{Comparison of patch precision on \swebenchlite. If \toolname does not use regression tests for patch validation, precision reduces only slightly to 48.3\%.}
    \label{fig:precision}
\end{figure}

A metric of particular interest to program improvement techniques is the precision. The precision is defined as TP/(TP+FP), i.e., the proportion of correct patches in all the patches offered by a technique. It is directly related to user effort required to examine generated patches. For \toolname, the precision is 50.0\%, as calculated in Table~\ref{tab:precision}. To put the precision in perspective, we compare the precision of \toolname with that of other baseline tools in Figure~\ref{fig:precision}. The precision of the baseline tools is the same as their pass@1 efficacy reported in Table~\ref{tab:efficacy}, since these tools indiscriminately present every generated patch to a user. As can be seen in Figure~\ref{fig:precision}, the precision of \toolname is higher than 1.8x that of Agentless, which has the second highest precision. The high precision of \toolname indicates a much lower cognitive load imposed on the user, compared to other techniques. Moving forward, we suggest paying attention to agent precision.


\subsection{RQ3: Quality of Specifications produced}

\begin{figure}[tb]
    \centering
    \begin{subfigure}[b]{\columnwidth}
        \centering
        \includegraphics[trim=0mm 192.5mm 79.5mm 3mm, clip,width=0.8\columnwidth]{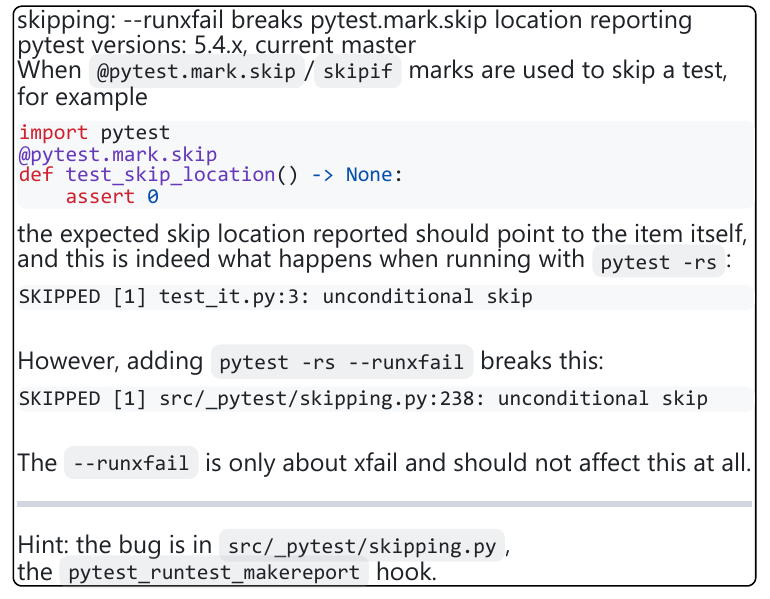}
        \vspace{-5pt}
        \caption{Issue statement.}
        \label{fig:pytet-7432}
    \end{subfigure}

    \vspace{5pt}
    
    \begin{subfigure}[b]{\columnwidth}
        \centering
\includegraphics[trim=0mm 250mm 81mm 2mm, clip,width=0.8\columnwidth]{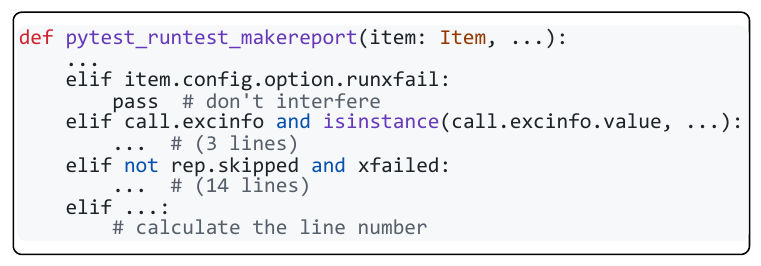}
        \vspace{-5pt}
        \caption{Buggy code snippet.}
        \label{fig:pytet-7432-bug}
    \end{subfigure}

    \vspace{5pt}

    \begin{subfigure}[b]{\columnwidth}
        \centering
        \includegraphics[trim=0mm 239mm 67mm 2mm, clip,width=0.8\columnwidth]{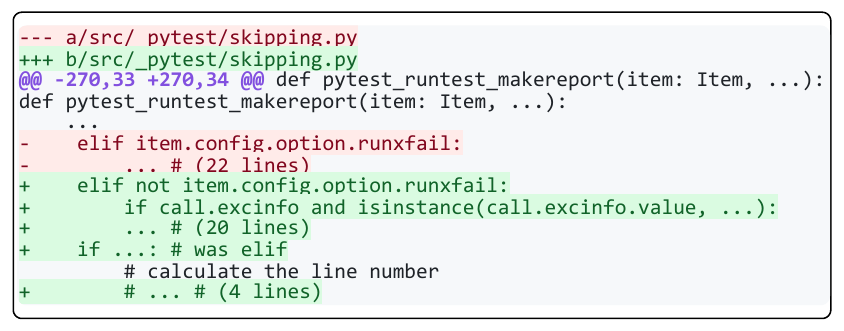}
        \vspace{-5pt}
        \caption{Correct patch generated by \toolname.}
        \label{fig:pytet-7432-patch}
    \end{subfigure}

    \vspace{5pt}

    \begin{subfigure}[b]{\columnwidth}
        \centering
        \includegraphics[trim=0mm 270mm 71.5mm 2mm, clip,width=0.8\columnwidth]{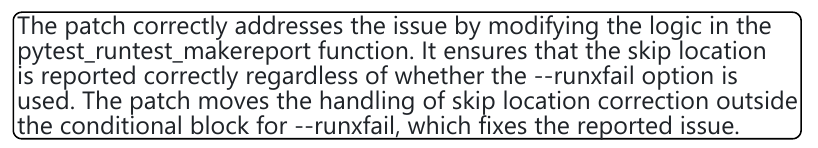}
        \vspace{-5pt}
        \caption{Reviewer feedback.}
        \label{fig:pytest-7432-feedback}
    \end{subfigure}

    \caption{pytest-7432 description and \toolname artifacts.}
\end{figure}

\begin{figure}[tb]
\begin{subfigure}[t]{0.47\columnwidth}
    \centering
    
    \includegraphics[trim=0in 9.94in 5.13in 0in, clip,width=\columnwidth]{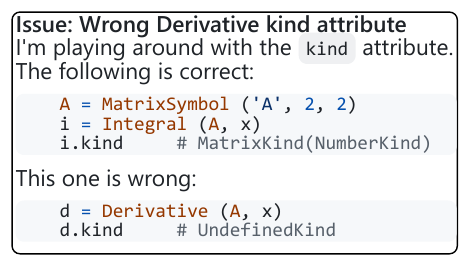}
    \vspace{-18pt}
    \caption{Issue statement.}
    \label{fig:sympy-21614-stmt}
\end{subfigure}
~
\begin{subfigure}[t]{0.47\columnwidth}
    \centering
    \includegraphics[trim=0.4in 9.58in 4.82in 0.4in, clip,width=\columnwidth]{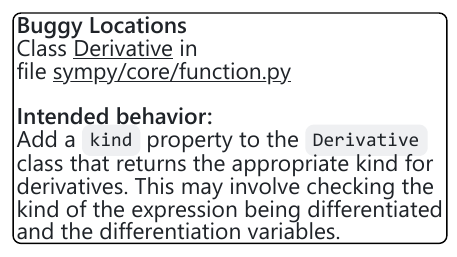}
    \vspace{-18pt}
    \caption{Location and summary.}
    \label{fig:sympy-21614-location}
\end{subfigure}

\vspace{5pt}

\begin{subfigure}{\columnwidth}
    \centering
    \includegraphics[trim=0.4in 9.39in 3.79in 0.4in, clip,width=0.8\columnwidth]{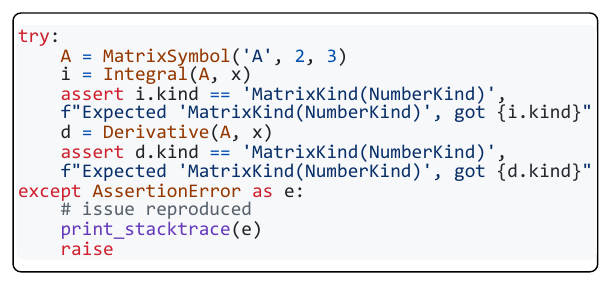}

    \vspace{-8pt}
    \caption{Generated reproducer test.}
    \label{fig:sympy-21614-reproducer}
\end{subfigure}

\vspace{5pt}

\begin{subfigure}{\columnwidth}
  \centering
  \includegraphics[trim=0.42in 9.79in 4.1in 0.45in, clip,width=0.8\columnwidth]{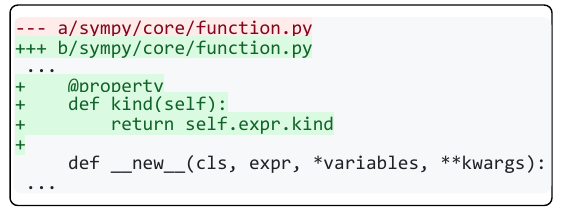}
  
  \vspace{-7pt}
  \caption{Correct patch generated by \toolname.}
  \label{fig:sympy-21614-patch}
\end{subfigure}

\vspace{5pt}

\begin{subfigure}{\columnwidth}
    \centering
    \includegraphics[trim=0.4in 9.86in 3in 0.4in, clip,width=0.8\columnwidth]{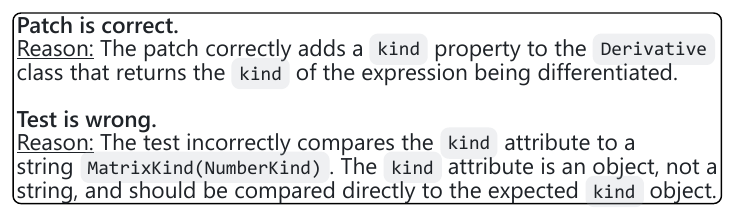}
    \vspace{-6pt}
    \caption{Reviewer feedback.}
    \label{fig:sympy-21614-review}
\end{subfigure}



  

\caption{SymPy-21614 description and \toolname artifacts.}
\label{fig:sympy-21614}
\end{figure}

In this section, we \new{investigate} the quality of evidence generated by \toolname. The high-quality evidence allows a developer to easily integrate auto-generated patch into an existing codebase.

\subsubsection{\new{Function Summaries}}
\new{
We manually investigated the natural-language specifications generated by \toolname for \swebenchlite issues. To obtain some approximation of ``ground truth'' for the generated specifications, we extracted the titles of Pull Requests that developers wrote when they fixed each issue. The PR title usually summarizes the fixes made in the PR in one sentence. In this investigation, we include only the issues for which \toolname generated a patch at the same methods as the developer's patch. Moreover, we exclude the issues whose developer-written PR title is too brief (e.g. ``Fixes issue \#...''). For the included issues, we manually compare the specification generated by \toolname for the patched methods, with the PR title written by the developer. For 72/101 (71\%) inspected issues, the specification generated by \toolname covers similar intent to the human-written PR title, which shows that the generated specification serves as a good summary of the intended behavior. The generated specifications are generally an extended version of the PR title and additionally describe how the code should be modified.

}

\subsubsection{\new{Reviewer Feedback}}

\new{We cannot conduct a quantitative examination on the quality of the reviewer feedbacks on the generated patches, since there is no ``ground truth''. Therefore, we illustrate the quality of the feedbacks with two examples.}

\paragraph{Reviewer Feedback as Summary}
In the first example, we show that the reviewer feedback can serve as a concise summary of a generated patch. 
The summary describes the behavior of the patch at a high level. 
Therefore, a developer can understand the generated patch faster by reading the summary before examining the details of the patch. 
Besides, after the developer accepts the patch and decides to merge it into existing code, the summary constitutes a good commit message, so that the developer does not need to write one. 
From a developer's perspective, the whole process is very much like reviewing a pull request, which is already part of a developer's day-to-day workflow. 
The specific issue involved in this example is pytest-7432\footnote{\url{https://github.com/pytest-dev/pytest/issues/7392}}. 
The issue statement is shown in Figure~\ref{fig:pytet-7432}, which reports that pytest (a python testing framework) would miscalculate a line number in its output when an irrelevant option (\code{runxfail}) is enabled. The bug is caused by the code shown in Figure~\ref{fig:pytet-7432-bug}. As can be seen, the calculation of the line number is wrongly placed in a branch that is mutaully exclusive with the \code{runxfail} branch. Therefore, the calculation is wrongly skipped when the option is enabled.

To resolve the issue, \toolname was able to locate the relevant code and produce the correct patch. An abridged version of the patch is shown in Figure~\ref{fig:pytet-7432-patch}. It correctly addresses the issue by moving the line number calculation to a branch unaffected by the \code{runxfail} option. However, the patch might not be immediately understandable to a developer, because it changes as many as 51 lines in the original program (though most of the changes just involve the indentation level).
Fortunately, the understanding of the patch can be eased by the reviewer feedback. The reviewer agent was able to identify the patch as correct and produced the feedback in Figure~\ref{fig:pytest-7432-feedback}. It  properly summarized that the patch just moved the calculation to another branch. Using this summary, the developer would easily understand the patch and accept it.

\paragraph{Tolerance of Incorrect Tests}

Another advantage of the reviewer is enhanced tolerance of incorrect automatically generated tests. We illustrate this advantage with the example of SymPy-21614\footnote{\url{https://github.com/sympy/sympy/issues/21604}}, where \toolname rejects an incorrectly written test while approving a correct patch.
The issue statement and the buggy location identified by \toolname are shown in Figure~\ref{fig:sympy-21614-stmt} and \ref{fig:sympy-21614-location}.
The issue mentioned an unexpected behavior of the \texttt{kind} attribute.
After its context retrieval stage, \toolname correctly identifies that the buggy location is in the \texttt{Derivative} class, and that its intended behavior is to have an additional \texttt{kind} property.
Figure~\ref{fig:sympy-21614-reproducer} and \ref{fig:sympy-21614-patch} show the automatically generated reproducer test and patch for this issue. 
In this case, the reproducer test is incorrect - the assertions compare an object with a string, which always evaluate to \texttt{False}.
If this reproducer test is used solely to determine the correctness of the generated patches, any patch, even a correct one, will be rejected.
However, since the reviewer agent in \toolname simultaneously examines both the reproducer test and the patch without assuming either is correct, it is capable of rejecting the reproducer test while approving the patch.
Figure~\ref{fig:sympy-21614-review} shows the reviewer agent's decision and comments towards the test and patch.
The reviewer identifies that the assertions in the reproducer are written incorrectly, thereby rejecting the reproducer.
On the other hand, the reviewer correctly approves the patch despite the presence of an unreliable test. 
The correct patch, along with the reasons for rejecting the reproducer, can be sent to the developer. 
The developer can then integrate the patch into the codebase. 
Additionally, the developer can revise the ``almost correct" reproducer test based on the feedback provided by the reviewer agent. 





\begin{figure}[tb]

	\begin{subfigure}{\linewidth}
		\centering
		\includegraphics[trim=0 253.5mm 75.5mm 2mm, clip,width=0.9\columnwidth]{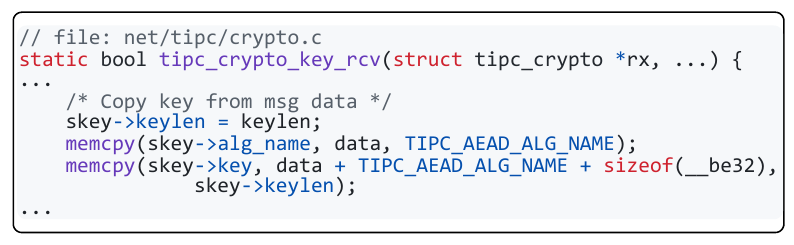}

		\vspace{-5pt}
		\caption{Buggy code snippet.}
		\label{fig:linux-bug}
	\end{subfigure}

    \vspace{5pt}

	\begin{subfigure}{\linewidth}
		\centering
		\includegraphics[trim=0.4in 9.65in 3.01in 0.4in, clip,width=0.9\columnwidth]{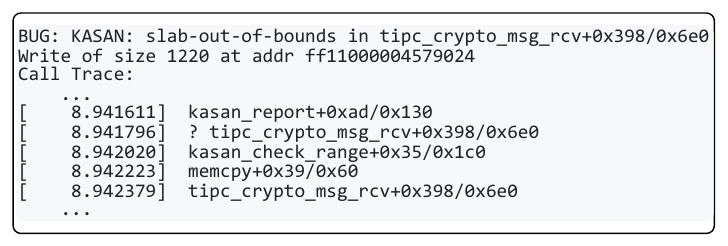}
		\vspace{-5pt}
		\caption{Vulnerability report generated after executing the PoC.}
		\label{fig:linux-report}
	\end{subfigure}

    \vspace{6pt}

	\begin{subfigure}{\linewidth}
		\centering
		\includegraphics[trim=0 243.5mm 86mm 2mm, clip,width=0.9\columnwidth]{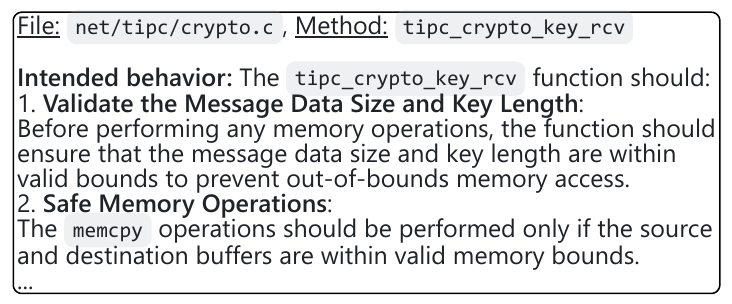}

		\vspace{-5pt}
		\caption{Buggy location and function summary by \toolname.}
		\label{fig:linux-locations}
	\end{subfigure}

    \vspace{5pt}

	\begin{subfigure}{\linewidth}
		\centering
		\includegraphics[trim=0 206.5mm 76.5mm 2mm, clip,width=0.9\columnwidth]{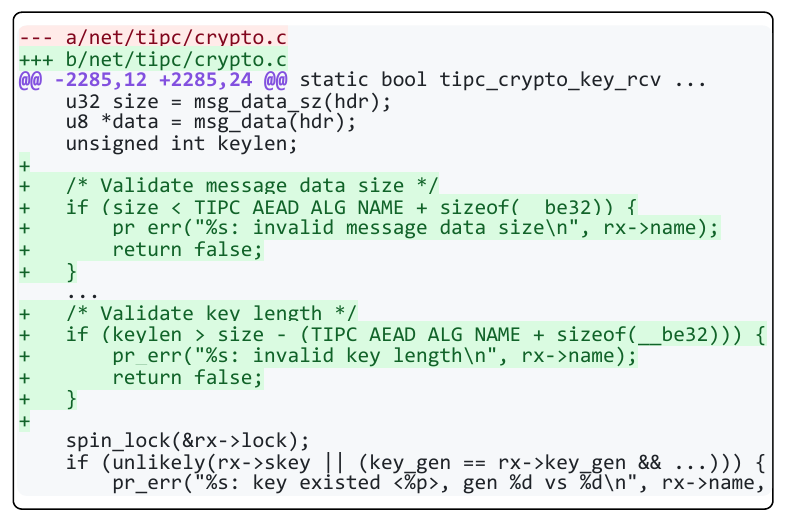}
		\vspace{-5pt}
		\caption{Correct patch generated by \toolname.}
		\label{fig:linux-patch}
	\end{subfigure}

    \vspace{5pt}

	\begin{subfigure}{\linewidth}
		\centering
		\includegraphics[trim=0 257mm 69mm 2mm, clip,width=0.9\columnwidth]{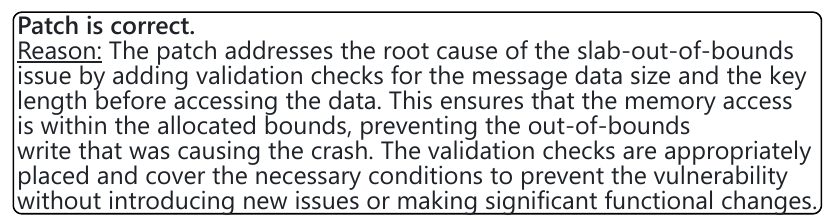}

		\vspace{-5pt}
		\caption{Reviewer feedback.}
		\label{fig:linux-review}
	\end{subfigure}

	\caption{Linux Kernel Exemplar Challenge in DARPA AIxCC.}
	\label{fig:linux}
\end{figure}

\section{Case Study: Security Vulnerability Repair}

Although \toolname is initially designed to resolve GitHub issues in Python repositories, it can be easily adapted for program improvements in other application domains, and for programs written in other programming language.
We demonstrate how \toolname fixes security vulnerabilities in C programs, through an example challenge problem from the DARPA AI Cyber Challenge (AIxCC) in 2024~\cite{aixcc}.
The AIxCC is a two-year competition organized by DARPA and ARPA-H to encourage the development of novel cyber-reasoning systems to safeguard critical software.
The AIxCC has publicly released exemplar challenges, where each challenge consists of a software project with a vulnerability in it.
The task is to have an autonomous system to find and fix the vulnerability.
Each exemplar challenge also contains a Proof-of-Concept (PoC) input file that triggers the vulnerability, so we use this PoC to show how \toolname can be used to fix the vulnerability after it is detected.
Figure~\ref{fig:linux} shows one exemplar challenge, which is a buffer overflow vulnerability in the Linux kernel\footnote{CVE-2021-43267 re-introduced to Linux kernel 6.1.54}. 
This buffer overflow happens in the Linux networking module for the Transparent Inter-Process Communication (TIPC) protocol, and allows remote attackers to cause denial-of-service or disclosure of sensitive information.
Specifically, when the user-supplied sizes in the message body are invalid for the received messages, a buffer overflow happens with the \texttt{memcpy} call, as shown in Figure~\ref{fig:linux-bug}.
This vulnerability has been triggered by a PoC, which results in a \textit{vulnerability report} as shown in Figure~\ref{fig:linux-report}.


\toolname fixes this vulnerability by first analyzing the vulnerability report, similar to how it resolves GitHub issues by initially examining the issue descriptions. 
It conducts context retrieval, and decides on the buggy locations and intended behaviors as shown in Figure~\ref{fig:linux-locations}.
Even though the vulnerability report only contains the call trace and minimal description of the bug (e.g., ``slab-out-of-bounds''), \toolname can infer the intended local behavior at the function level.
Based on the intended behavior, \toolname generated the patch in Figure~\ref{fig:linux-patch}, which correctly fixes the vulnerability inserting additional checks before the dangerous memory operation.
The reviewer agent approved the patch with the comments shown in Figure~\ref{fig:linux-review}, with which the developers can gain an initial understanding of the patch before closely examining the changed code.

\section{Related Work}

Automated program repair (APR) \cite{cacm19,monperrus} is a well studied research area in software engineering. Given a buggy program $P$, and a test-suite $T$, automated program repair attempts to (minimally) modify $P$ to a program $P'$ which passes the given test-suite $T$. APR techniques involve metaheuristic search \cite{genprog}, semantic analysis \cite{semfix}, machine learning \cite{prophet}, or a combination of different techniques. APR can also be used to rectify automatically generated code from LLMs, see e.g. \cite{Fan23}.

The recent interest in prompt engineering as well as agent based solutions has somewhat evolved the research in program repair. LLM agents try to combine the power of LLM with program analysis and test execution reasoning. Thus LLM agents can combine LLMs with test generation, static and dynamic analysis as well as specification inference. In the recent past, lot of LLM based approaches have been proposed for solving software ``issues'' described in natural language, including \cite{devin,autocoderover,swe-agent,agentless}. Among these our work is thematically closest to the work of AutoCodeRover \cite{autocoderover}. Like AutoCodeRover. we take the position that program modifications like bug fixing are best aided by inference of the developer intent. AutoCodeRover infers the developer intent only from the software project structure. In contrast, \toolname is more general and is capable of inferring specifications from different sources including program structure, program behavior, tests and so on. Furthermore, \toolname focuses on giving an explanation of the produced patches.

\new{Previous works have studied function-level specification inference by means of LLMs~\cite{mu2024clarifygpt,ma2024specgen,madeline2024nl2postcond}. However, these works either focus on generating specifications for simple one-function or one-class programs~\cite{ma2024specgen,mu2024clarifygpt}, or assume that a target function is provided~\cite{madeline2024nl2postcond}. In contrast, \toolname targets large programs and does not assume the target functions to generate specifications for are given.}

\new{Since both tests and patches generated in an autonomous workflow can be unreliable, we additionally consider the natural-language issue description when judging their correctness. The judgement process is concretized as the reviewer agent. Compared to the previous works 
that discover additional specification based on user intent~\cite{Fakhoury24interactive}, we do not rely on interactive user feedback. Instead, we utilize the high-level natural-language description as additional feedback.
Furthermore, compared to previous works that evaluate the generated code with an LLM-based reviewer~\cite{zhang23coderreviewer}, our reviewer agent is designed for the setup where both natural-language instructions and unreliable tests are present.
}

\section{Perspectives}
Owing to the growth of LLM-based automatic programming  (see \cite{llm-arxiv} for a recent summary), there exists interest in autonomous program improvement technologies. We propose \toolname with the perspective of autonomously producing patches which are suggested with confidence (thus developers can confidently accept them) and come with explanations. The technical innovations supporting \toolname are the specification inference to guide patching, and the rigorous vetting of patches via our reviewer agent. Our work on \toolname seeks to put the matter of quality of patches produced by LLM agents into the research community's attention, whereas other works are mostly focusing on the agent efficacy.

Moving forward, we envision that LLM agents will need to improve the precision and recall. Specifically, LLM agents will need to vet the produced patches. The vetting needs to be accompanied by sophisticated test generation, so that there is a rich test-suite to check the produced patches.  Thus, we need to take the viewpoint of a developer using an LLM agent, who would be concerned about (a) efficacy, (b) cost, and, most importantly (c) {\em signal-to-noise ratio} in the agent output.


\section*{Data Availability}

We share full public access of the source code and experimental artifacts of \toolname at \artifacturl.

\new{We are continuously improving the efficacy and usability of \toolname (i.e., \acrvtwo). At the time of paper acceptance (Nov 2024), \acrvtwo has achieved an efficacy of 37.3\% on \swelite and 46.2\% on SWE-bench Verified. We have also packaged \acrvtwo as a GitHub bot offering one-click issue resolution~\cite{acrbot}. The latest source code, experimental results, and news updates of \acrvtwo can be found at our GitHub repository\footnote{\url{https://github.com/nus-apr/auto-code-rover}} and website\footnote{\url{https://www.autocoderover.net/}}.

}

\section*{Acknowledgments}

\new{This work was partially supported by a Singapore Ministry of Education (MoE) Tier 3 grant "Automated Program Repair", MOE-MOET32021-0001.}

\bibliographystyle{IEEETran}
\bibliography{references}

\end{document}